**First-principles DFT study for the structural stability of Hydrogen passivated graphene (H-graphene) and atomic adsorption of oxygen on H-graphene with different schemes.**

D. B. Karki and N. P. Adhikari[*]
*Central Department of Physics, Tribhuvan University, Kirtipur, Kathmandu, Nepal*
[*]Corresponding author: npadhikari@tucdp.edu.np

**Abstract**

First-principles DFT levels of calculations have been carried out in order to study the structural stability and electronic properties of hydrogen passivated graphene (H-graphene) clusters. Two different shaped clusters, rectangular and circular, consisting of 6 to 160 carbon atoms and hydrogen termination at the zigzag boundary edges have been studied. The relative stability of circular shaped cluster consisting 96 C-atoms have been predicted to be around 1.5% greater than that of rectangular shape cluster consisting same number of C-atoms. In comparing circular and rectangular cluster containing same number of C-atoms, the HOMO-LUMO gap of former have been predicted to be 2.159 eV and that of later just 0.346 eV. Adsorption of oxygen atom on H-graphene with different schemes including single sided, both sided and high concentration adsorption, was also studied systematically through first-principles DFT calculations by taking four different H-graphene clusters. The calculations showed that the most stable adsorption site for oxygen adatom on H-graphene being B-site with adsorption energy 4.011 eV on the rectangular H-graphene cluster consisting 70 carbon atoms. Moreover, on increasing the size of H-graphene cluster, the adsorption energy of oxygen atom found to be increase. The distance of adatom from the nearest carbon atom of H-graphene sheet was 1.52 Å, however, the adatom height from the H-graphene basal plane was 1.97 Å. The bonding of oxygen adatom on H-graphene was through the charge transfer about 0.40 |e| from H-graphene to adatom and includes the negligible local distortion in the underlying planner H-graphene. Charge redistribution upon adsorption induces significant dipole moment 2.356 *Debye* on rectangular H-graphene cluster consisting 70 carbon atoms. The adsorption energy per O-atom in case of both side adsorption (one at B site and other at opposite B site below the sheet) was found to be around 2% greater than that of single O adsorption. The calculated values of dipole moment (0.881 *Debye*) and HOMO-LUMO gap (0.590 eV) in this case were almost one third of that for single O adsorption. The adsorption energy per O atom for both side adsorption model such that one at the B site and other at neighboring B site below the H-graphene sheet was found to be 4.650 eV, which is around 14% greater than that of both side adsorption discussed above and 16% greater than that of single O adsorption. The adsorption energy per O-atom, dipole moment and HOMO-LUMO gap in case of three oxygen atom adsorption on the alternate B site of central benzene ring of H-graphene cluster $C_{70}H_{22}$ have been estimated to be 4.522 eV, 5.898 *Debye* and 0.820 eV respectively.

**Keywords**

H-graphene, Density Functional Theory (DFT), Adsorption, charge transfer, dipole moment, HOMO-LUMO energy gap

## 1. Introduction

The fast growing technological importance of carbon-based nanostructures is well known [1]. From the backbones of organic molecules to particulates in air pollution [2, 3] or diamond-like films, carbon is crucial to the stability and properties of many natural and artificial structures. An infinite number of covalent carbon structures may exist with either diamond-like ($sp^3$), graphite-like ($sp^2$), linear (sp) or hybrid ($sp^3/sp^2$, $sp^2/sp$, $sp^3/sp$) bonding [4]. Graphene, the single layer of graphite or one-atom-thick planar sheet of $sp^2$ bonded carbon atoms tightly packed in a honeycomb crystal lattice is the basis of all the graphitic forms; 0D Fullerenes [5], 1D carbon nanotubes [6], and 3D graphite which exists in pencil. The advent of graphene [7] has excited much interest in these unique 2D systems because of the enormous potential that graphene has for novel electronic, magnetic and optical applications. Graphene has grabbed much attention due to its gapless electronic band structure with a pointlike Fermi surface and a linear dispersion at the Fermi level. These properties are responsible for the observed ballistic transport, Dirac-type quasiparticles, and anomalous quantum Hall effects in graphene [8]. More, recently, simple graphene based devices have become feasible, demonstrating that the initial concept purposed for graphene can be realized [9].

In literature one can hardly find little works, whose main point of research was for the stability of H-graphene clusters. Oli *et al* [10] treated both type of H-graphene clusters in the same footing without any signature for the shape dependent stability and other electronic properties. Likewise, in the work [11], they studied only the circular shape H-graphene clusters and in the work [12], we just studied the stability of rectangular shape H-graphene clusters. To our best knowledge no works has been reported for the comparison of electronic properties between circular and rectangular shaped H-graphene clusters. Thus, study for the shape dependent stability and electronic properties of H-graphene clusters kept immense value to tailor the properties of graphene. With this regard, we have made series of calculations to study the stability and other properties of circular and rectangular shaped H-graphene clusters.

The interaction of graphene with atoms is so important because of their fundamental relevance to applications, such as catalysis, batteries and transistors [13]. Thus, to study the mechanism of adatom adsorption by graphene is indispensable for understanding the growth of the compounds on graphene. In the past few years, studies of atomic adsorption on graphene have attracted considerable attentions [14-19], but these studies only conducted calculations for monoatomic adsorption. Furthermore, there were some attempts and calculations to study the adsorption on graphene with greater coverage of adatoms including both sided adsorption with different schemes [20, 21]. In our best knowledge, no studies have discussed the coverage change of oxygen atom on graphene including both sided adsorption with different schemes. Oxygen is one of the important materials in electronic devices [22], in as much graphene oxides is novel material for their possible use in solar cells, thermoelectric devices, and water filtration, among a number of other applications. Modifying graphene through the addition of oxygen atoms can provide those properties. With these considerations, in this research, we chose to study the adsorption on H-graphene with oxygen atoms as the adatoms.

In the present work, we study for the structural stability of hydrogen passivated graphene (H-graphene) taking rectangular shaped and circular shaped H-graphene clusters consisting of 6 to 160 carbon atoms and hydrogen termination at the boundary edges performing hybrid functional calculations employing the popular Becke-three-parameters-Lee-Yang-Parr hybrid functional [23, 24] together with the basis set 3-21G. Atomic adsorption of oxygen atom on H-graphene clusters considering different schemes have been also studied. We calculated the adsorption energies and electronic properties of O-adsorbed H-graphene with different schemes, including one sided adsorption, both sided adsorption and high concentration adsorption. These research results would help researchers to tailor the properties of graphene such as the H-L gap specifiy metal-semiconductor properties. The relative stability specifies the clusters which show a better performance at higher temperatures. Moreover, by extrapolation of data, extra detail of the behavior of graphene can be achieved.

This paper is organized as follows: in section 2, we briefly describe the method of the calculations. In section 3, we present the results and discussion of the present work and conclusion will be found in section 4.

## 2. Methodology

We have performed first-principles calculation based on Density Functional Theory (DFT) [25, 26] with Gaussian orbital functions for the basis sets. With DFT, the gradient corrected functional B3LYP (i.e. Becke's 3-parameter hybrid exchange functional [19] and Lee, Yang, and Parr correlation functional [20]) has been employed for our calculations. The electronic structures and total energies are calculated using Gaussian 03 [27] set of programs with the choice of basis set 3-21G.

For the study of the stability of H-graphene clusters, each cluster was first optimized without any symmetry restriction using Gaussian 03 set of programs with the choice of basis set 3-21G. To ensure the global minima, each optimize geometry was used to run the frequency calculation. For each cluster, all the calculated frequencies were real, confirming each optimize structure with global minima. The binding energy of H-graphene clusters have been calculated by using the relation; $E_B = N_C E_C + N_H E_H - E_{H\text{-}graphene}$ where, $N_C$ and $N_H$ are the number of C-atoms and H-atoms in the H-graphene cluster, $E_C$ and $E_H$ are the ground state energies of isolated C-atom and H-atom respectively and $E_{H\text{-}graphene}$ is the ground state energy of corresponding H-graphene cluster. The binding energy per atom has been defined to be the binding energy of H-graphene sheet divided by corresponding total number of atom ($N_C + N_H$).

We have consider the binding of oxygen atom on three sites of high symmetry as shown in Fig. 1; the hollow (H) site at the centre of a hexagon, the bridge (B) site at the midpoint of carbon-carbon bond, and the top (T) site directly above the carbon atom. For each adsorption site, the adatom is relaxed along the Z-direction considering H-graphene sheet in XY-plane. The minimum energy of H-graphene systems with oxygen atom adsorption along B-site shows that adsorption of oxygen atom on H-graphene is most favourable in the bridge (B) position agreeing with the adsorption of oxygen atom on graphene [28] and hence in the present work we discuss only the adsorption in the B-site. The optimization of the adatom-H-graphene system is performed in the following steps. First, we optimized the geometries of H-graphene clusters. After that, the position of the adatom on H-graphene sheet at the bridge site is fixed above certain height and the distances between adatom and H-graphene sheet have been set to values that were slightly smaller than the sum of the adsorbate and carbon atoms covalent radii. And the final geometries of the adatom-H-graphene systems have been optimized. The binding (adsorption) energy of adatom on H-graphene sheet has been estimated as;

$$E_B = E_1 + E_2 - E_{1-2}$$

where $E_1$ is the ground state energy of adatom, $E_2$ is the ground energy of H-grapheme sheet and $E_{1-2}$ is that of the adatom adsorbed H-graphene sheet.

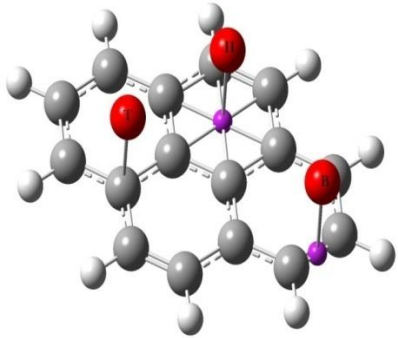

Fig.1 Different adsorption sites [Hollo site (H), Bridge site (B) and Top site (T)] of high symmetry for adsorption on H-graphene.

## 3. Results and Discussion

### 3.1 Rectangular H-graphene clusters

#### 3.1.1 Stability

We have performed the optimization calculations for eight different rectangular shaped H-graphene clusters; $C_6H_6$, $C_{16}H_{10}$, $C_{30}H_{14}$, $C_{48}H_{18}$, $C_{70}H_{22}$, $C_{96}H_{26}$, $C_{126}H_{30}$ and $C_{160}H_{34}$ using first-principles DFT(B3LYP) levels of approximation with the choice of basis set 3-21G implimented by Gaussian 03 set of programs. The optimized geometry of each studied cluster are as shown in Fig. 1, with the calculated C-C bond lengths in the range of 1.35 Å to 1.43 Å, some of these bond lengths are comparable with 1.42 Å characteristic of $sp^2$ carbon in graphene and some are too smaller. The C-H bond lenghts are 1.10 Å. To study the stability of H-graphene, we have calculated the binding energies of studied clusters using reference energies -1022.412 eV for C-atom and -13.533 eV for H-atom obtained using same calculations as for H-graphene clusters. The variation of calculated binding energy per atom (δE/N) for H-graphene clusters as a function of corresponding total number of atom (N) is shown in Fig. 2. From the Fig. 2, it is seen that δE/N increases with increase in cluster size and tending toward saturation. In other word, the larger size H-graphene clusters are relatively more stable than smaller sized. For the cluster $C_{160}H_{34}$ (which almost shows the bulk behaviour), the calculated value of binding energy 7.869 eV/atom is comparable with the previously reported values 7.37 eV/atom [29] for solid carbon, 8.03 eV/atom [10] for graphene sheet having 32 carbon atoms and 7.91 eV/atom [30] for 4x4 unit cell of graphene.

The C:H ratio for $C_6H_6$ is 1:1, however, with increasing cluster size the percentage fraction of H-atom decrease and that for C-atom increase. On the other hand, from Fig. 2 it is already seen that with increasing the cluster size, δE/N goes on increasing. In this work, we have made an effort to find out the exact relationship between δE/N and percentage fraction of H-atom/C-atom in H-graphene clusters. From our calculations, we observed that on increasing the cluster size, δE/N follows perfect linearity with percentage fraction of H-atom as well as C-atom, however, it increases linearly with percentage fraction of C-atom and decreases linearly with percentage fraction of H-atom in the H-graphene clusters. The best fitted lines for δE/N as a function of corresponding percentage fraction of H-atom and C-atom in H-graphene clusters are given as, δE/N = [-0.064 × (% fraction of H-atom) + 8.980] eV/atom and δE/N = [0.064 × (% fraction of C-atom) + 2.600] eV/atom respectively as shown in Fig. 3. Using either relation, we can predict the binding energy per atom for any H-graphene clusters regardless of its size.

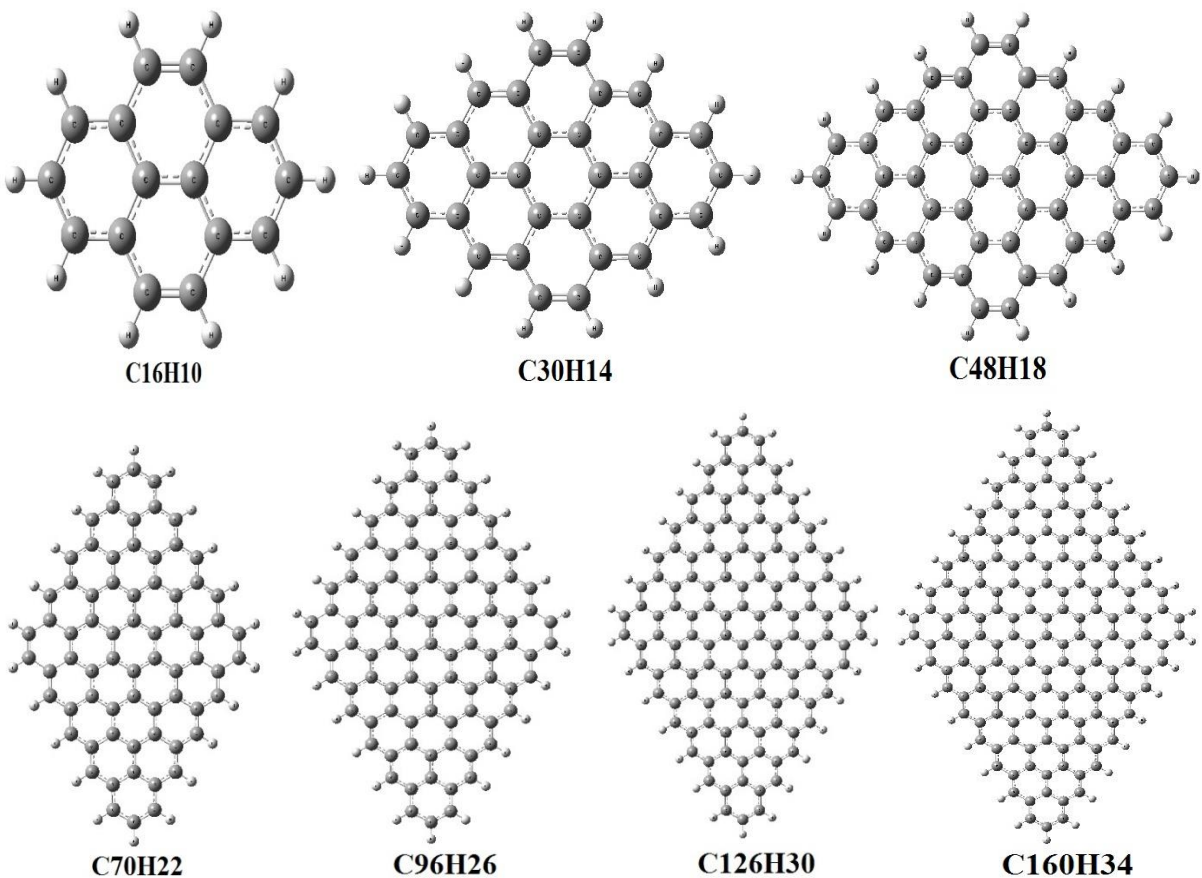

Fig. 1 Optimized geometries of rectangular shaped H-graphene clusters, where each carbon atom on the edge of the clusters has been passivated with Hydrogen atom and then were subjected to optimization using DFT(B3LYP) levels of approximation with the basis set 3-21G. For each clusters, the Mulliken charge amount (positive or negative) of each atom increases if that atom is nearer to the edge; also, hydrogen atoms on the edge have positive charges in the range of 0.185 |e| to 0.190 |e|.

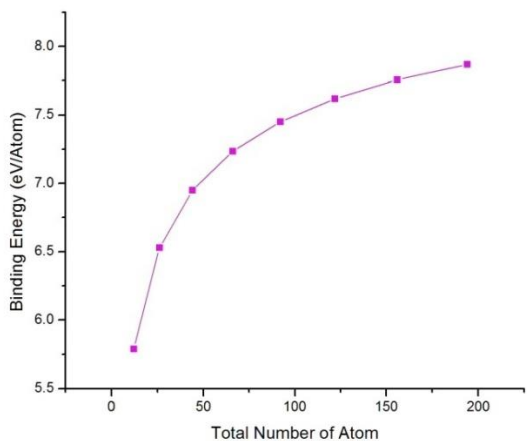

Fig. 2 Variation of the binding energy per atom in H-graphene clusters as a function of corresponding number of atoms ($N_C + N_H$) in DFT(B3LYP) levels of calculation with the choice of basis set 3-21G.

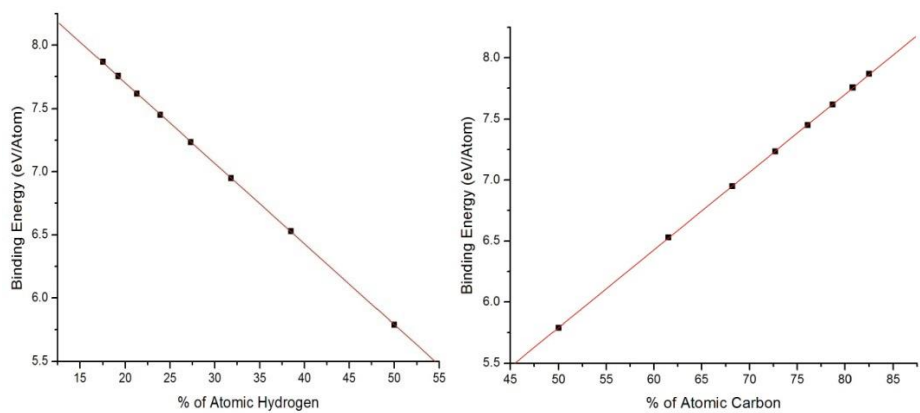

Fig. 3 The best fitted lines for the variation of binding energy per atom as a function of corresponding percentage fraction of H-atom and C-atom in H-graphene clusters. The left fitted line is δE/N = [-0.064 × (% fraction of H-atom) + 8.980] eV/atom and right one is δE/N = [0.064 × (% fraction of C-atom) + 2.600] eV/atom.

### 3.1.2 HOMO-LUMO (H-L) Energy Gap

Molecular orbitals (HOMO and LUMO) and their properties such as energy are very useful for physicist and chemist and are very important parameters for quantum chemistry. This is also used by the frontier electron density for predicting the most reactive position in $\pi$-electron systems and also explains several types reaction in conjugated systems [31]. The conjugated molecules are characterized by small HOMO, LUMO seperation, which is the result of significant degree of charge transfer form end-caping electron-donor groups to the efficient electron acceptor groups through $\pi$-conjugated path. The HOMO represents an ability to donate an electron (Ionization potential), LUMO as an electron acceptor, represents the ability to obtain an electron (Electron afinity).

The central aspect of the electronic properties is the energy gap between occupied and unoccupied electron states, the H-L gap. In the present work, we have systematically studied the structural dependence of H-L gap among H-graphene clusters using Natural Bond Orbital analysis (NBOs). The variation of HOMO, LUMO energies and corresponding energy gap as a function of cluster size is as shown in Fig. 4. From the Fig. 4, it is seen that the H-L energy gap decrease with increase in system size agreeing with the previous calculations [32]. In small H-graphene clusters quantum effect become important so large energy gap has appear. With increase in cluster size, the quantum effect become less important and thus larger size H-graphene clusters would have insignificant energy gap. From this analysis we can say that, for infinitely large H-graphene cluster the H-L energy gap would be zero (equivalently to say with zero band gap) and its electronic properties would be same as that of pure graphene.

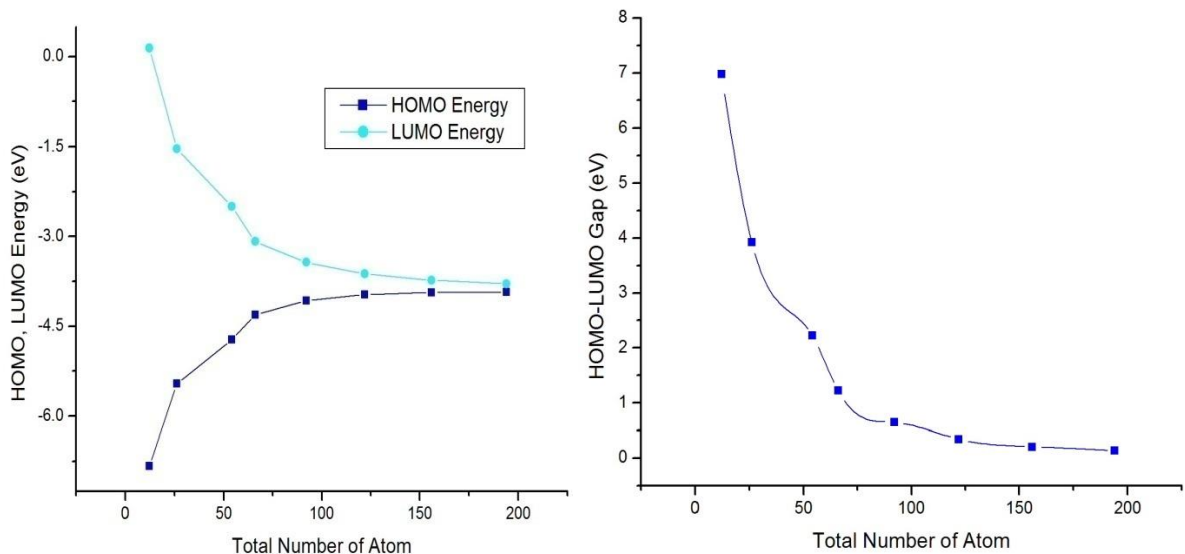

Fig. 4 Structural dependence of HOMO energy, LUMO energy and H-L gap among H-graphene clusters

The occupied and unoccupied molecular orbitals can be seen on density of states spectrum. Using Mulliken population analysis, we have also plotted the density of states (DOS) spectrum for each H-graphene cluster using Gauss sum 3.0 software [33] as shown in Fig. 5. From the figure, it is seen that, with increase in cluster size, the electron levels becomes closer and at the same time fermi level energy increases. Also it is seen that, the electron levels are not spin polarized and none of the clusters would have dipole moment.

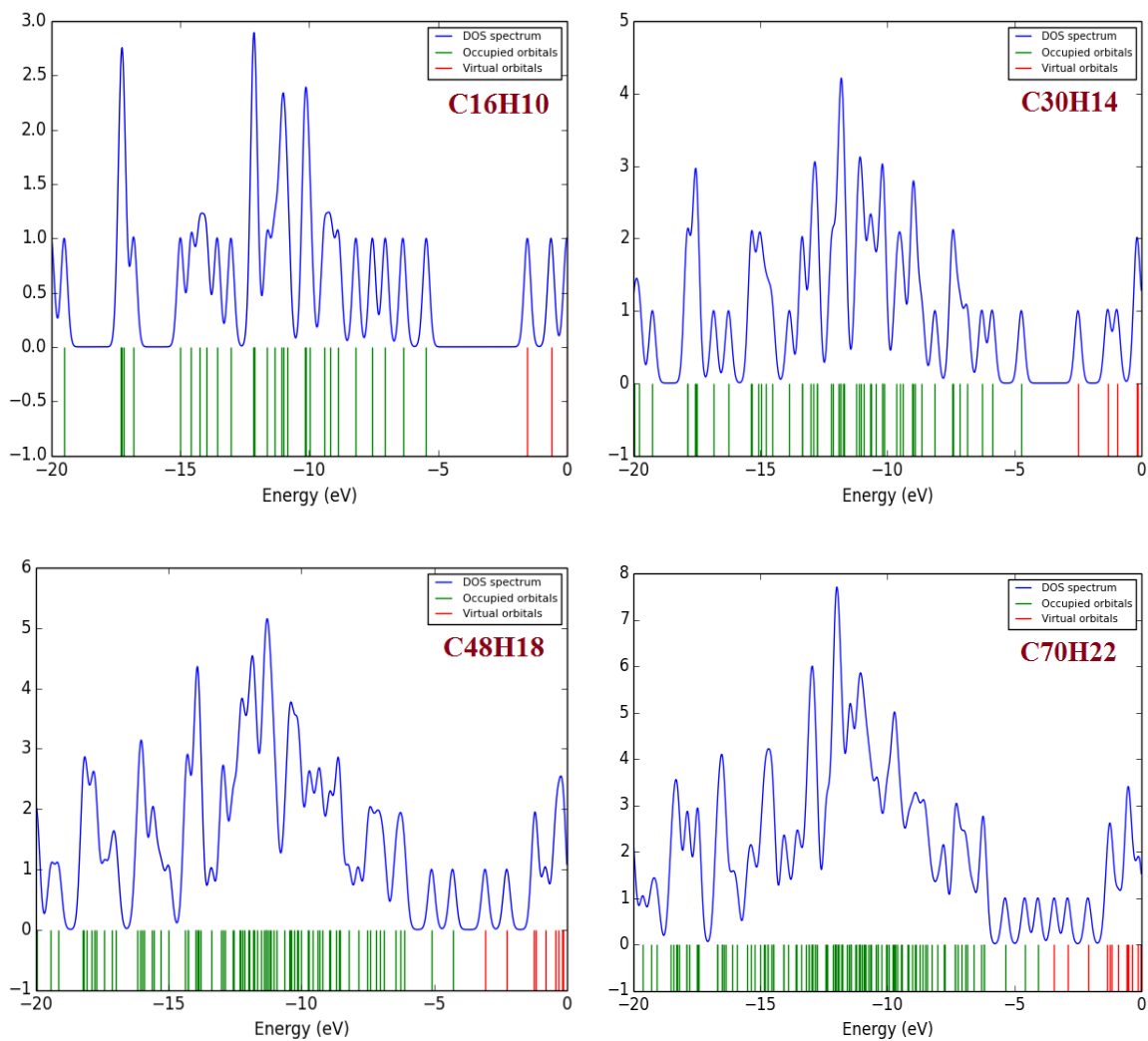

Fig. 5 DOS spectrum for some studied H-graphene clusters using Mulliken population analysis with DFT(B3LYP)/3-21G levels of calculation.

### 3.2 Circular H-graphene clusters

Same calculations as in rectangular shaped H-graphene clusters have been performed for four different circular shaped H-graphene clusters; $C_{24}H_{12}$, $C_{54}H_{18}$, $C_{96}H_{24}$ and $C_{150}H_{30}$ as shown in Fig. 6. The variation of calculated binding energy per atom as a function of corresponding total number of atoms in H-graphene clusters is as shown in Fig. 7. As in rectangular shaped H-graphene clusters, relative stability of circular shaped H-graphene clusters also increases with increase in cluster size and tends toward saturation. For the cluster $C_{96}H_{24}$, the calculated value of binding energy is 7.733 eV/atom. The corresponding value for rectangular shaped cluster containing same number of atoms, $C_{96}H_{26}$ is 7.620 eV/atom, showing that the relative stability of rectangular shape H-graphene clusters is lower than that of circular shape.

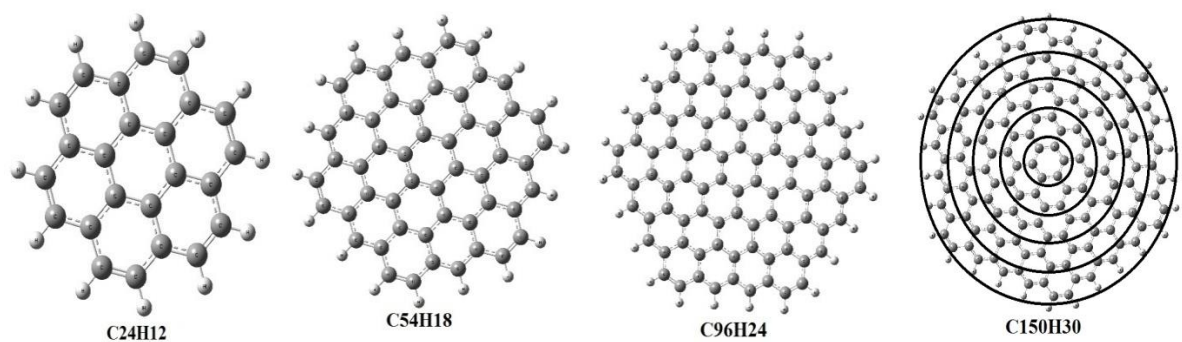

Fig. 6 Optimized geometries of circular shaped H-graphene clusters in DFT(B3LYP) levels of calculation with the choice of basis set 3-21G. The circling for $C_{150}H_{30}$ is to identify the number of rings. The H-graphene clusters; $C_{150}H_{30}$, $C_{196}H_{24}$, $C_{54}H_{18}$ and $C_{24}H_{12}$ has five, four, three and two benzene rings respectively.

For each four clusters, the charge amount (positive or negative) of each atom increases if that atom is nearer to the edge; also, each hydrogen atoms have almost equal positive charge 0.190 |e|. For $C_{150}H_{30}$, each C-atoms in the central and second ring has the NBOs charge of -0.002 |e| and -0.003 |e| respectively. The C-atoms in the third and fourth ring has the charges in the range -0.004 |e| to -0.010 |e|. For the C-atoms in the last ring, those which are connected with hydrogen atom has negative charges in the range of -0.170 |e| to -0.195 |e|, however, those which are not connected with hydrogen atom has the positive charges in the range of 0.016 |e| to 0.024 |e|. The calculated values of HOMO, LUMO energies and corresponding energy gap has been plotted as a function of total number of atom in the clusters as shown in Fig. 8 and found to be decrease with increasing cluster size. The decreasing pattern of H-L gap on increasing cluster size can also be seen from the corresponding DOS spectrum as shown in Fig. 9. In comparing circular and rectangular cluster containing same number of C-atoms, $C_{96}H_{24}$ and $C_{96}H_{22}$, the H-L gap of former is 2.159 eV and that of later is just 0.346 eV. Similarly, the H-L gap of $C_{150}H_{30}$ is 1.524 eV greater than that of $C_{160}H_{34}$. Thus, we can say that, relatively larger size circular H-graphene cluster would required to have zero gap than rectangular one.

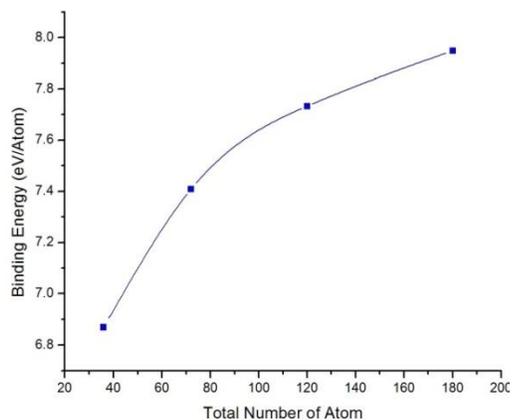

Fig. 7 Variation of the binding energy per atom in circular H-graphene clusters with corresponding number of atoms in DFT(B3LYP) level of calculations with the choice of basis set 3-21G.

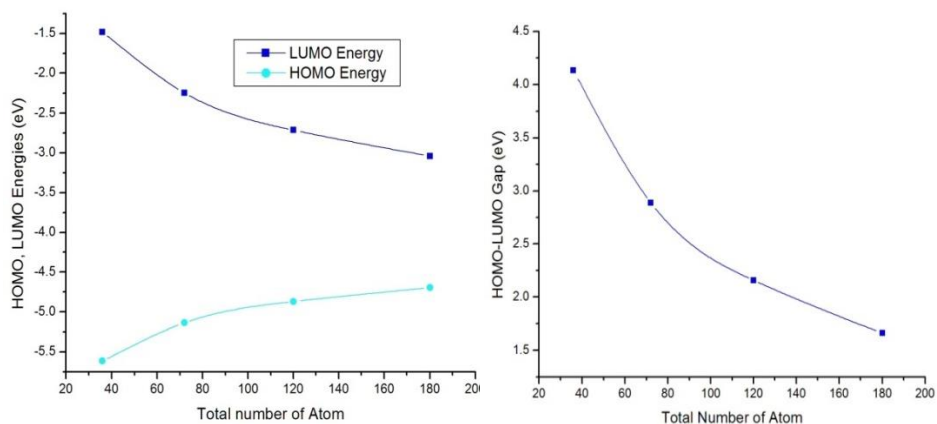

Fig. 8 Structural dependence of HOMO and LUMO energy, and H-L gap among circular shaped H-graphene clusters.

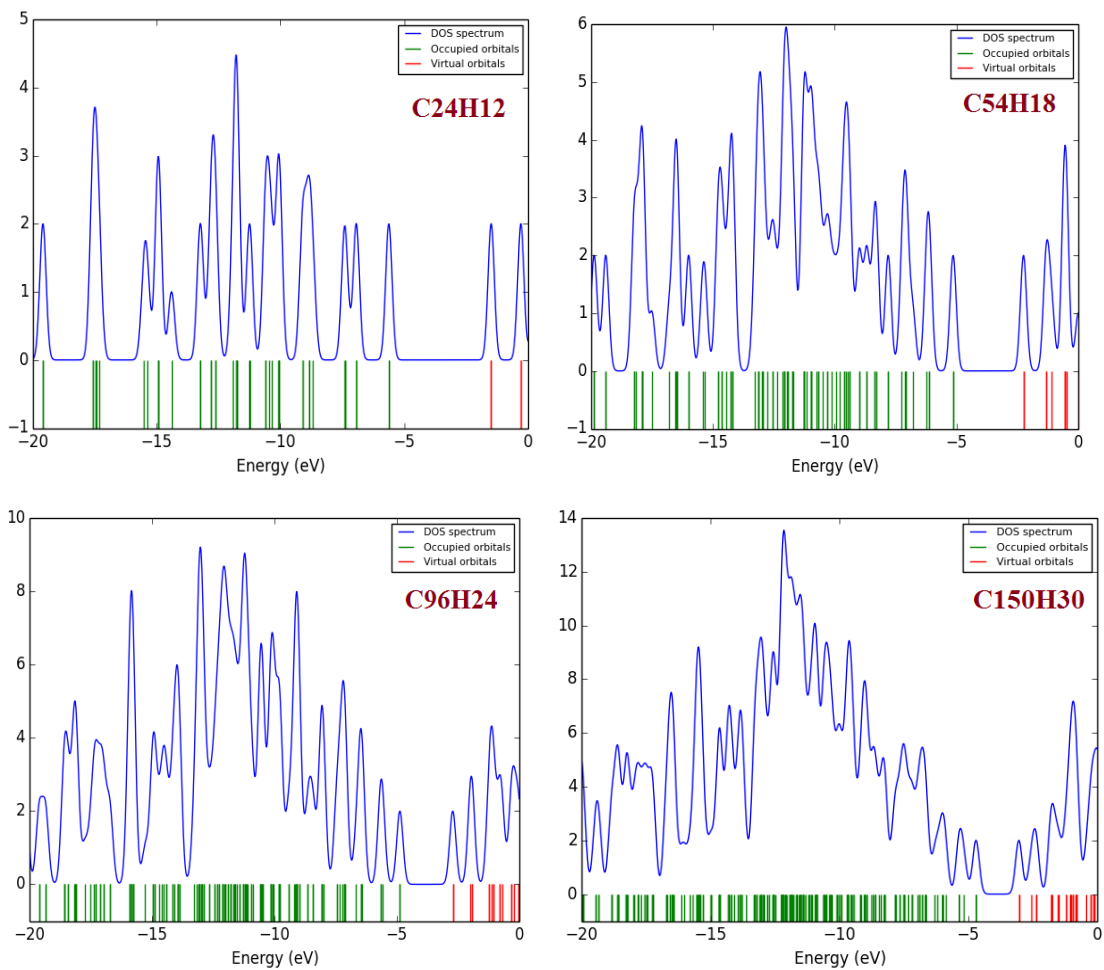

Fig. 9 DOS spectrum for circular H-graphene clusters using Mulliken population analysis with DFT(B3LYP)/3-21G levels of calculation.

It is evident that, for infinitely large H-graphene clusters, H-L gap would be zero. Previous study [11], shows that, the circular H-graphene cluster with about 500 C-atoms would have zero H-L gap. The circular H-graphene cluster with about 500 C-atoms is $C_{486}H_{54}$, which contain exactly 10 % of H-atoms. But, our study shows that, the circular H-graphene cluster with about 6 % H-atoms would have zero H-L gap. This can be seen from the best fitted line; H-L gap = $[0.1482 \times (\% \text{ of H-atoms}) - 0.8092]$ eV as shown in Fig. 10.

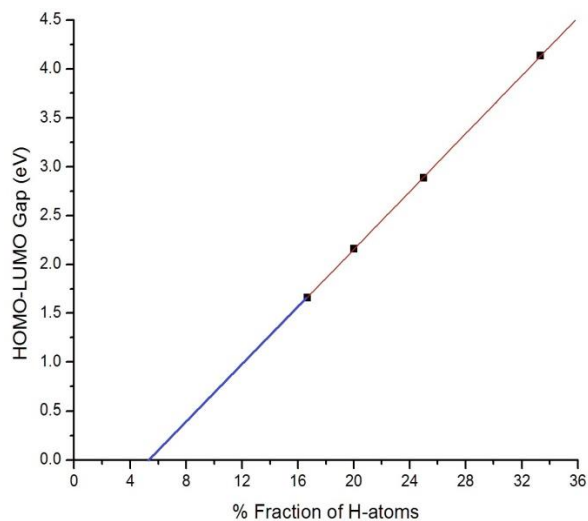

Fig. 10 Best fitted line between the percentage fraction of H-atoms in the H-graphene clusters and corresponding H-L gap. The red line is; H-L gap = [0.1482 × (% of H-atoms) – 0.8092] eV and the blue one is its extension showing that, the circular H-graphene cluster containing 5.46 % of H-atoms would have zero H-L gap.

### 3.3 Atomic adsorption of oxygen on H-graphene
### 3.3.1 Single-O adsorption

We investigate the adsorption of oxygen atom on flat H-graphene clusters. For this study, we have selected four H-graphene clusters; $C_{24}H_{12}$, $C_{30}H_{14}$, $C_{48}H_{18}$ and $C_{70}H_{22}$. The oxygen atom was kept at the distance of 1.60 Å above the H-graphene surface at the bridge (B) site, the most stable adsorption site for oxygen atom adsorption on H-graphene. The optimized structure of oxygen adatom adsorbed on $C_{70}H_{22}$ cluster is as shown in Fig. 11. From the view of this local structure of B-site adsorption, the distances of adatom from first-neighboring C-atom (1) and second- neighboring C-atom (2) were both about 1.52 Å, which is closed to the previously reported value 1.59 Å [34]. The adsorption of oxygen atom on each chosen cluster causes to change in bond lengths, bond angles and dihedral angles between C-atoms of H-graphene clusters and found to be decrease with increasing the cluster size. For the cluster $C_{70}H_{22}$, oxygen adsorption causes the maximum change in bond length 0.11 Å, bond angle 2.0º and dihedral angle 19.3 Å producing slight distortion (buckling) on the underline planner H-graphene sheet. The buckling is due to the fact that the adatom pulls the C-atoms away from the H-graphene layer. The bond lengths, angles and dihedrals, which are farther from the oxygen atom remain almost unchanged, however, the maximum change in these parameters can be seen from neighboring C-atoms of adsorbed oxygen.

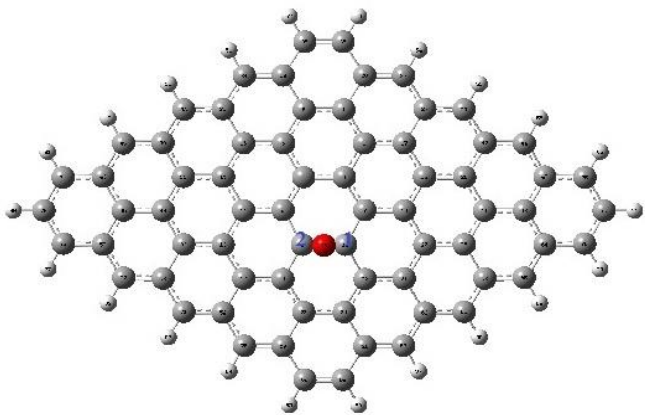

Fig. 11 Optimized structure of $C_{70}H_{22}$ H-graphene cluster with oxygen atom adsorbed at B site obtained with DFT(B3LYP) levels of calculation using the basis set 3-21G. Where, 1 and 2 are the first and second neighbors of adsorbed O-atom, on the bridge formed by these two C-atoms, the oxygen atom has adsorbed.

As stated earlier, none of the H-graphene clusters would have the dipole momentum, however, the adsorption of oxygen causes the charge redistribution producing net dipole moment about 2.3 *Debye* in each H-graphene cluster. This redistribution of charge also causes to change in H-L energy gap in H-graphene clusters and the H-L gap found to be decrease with increasing the cluster size. Our calculation shows that, the charge transfer of around 0.39 |e| takes place from H-graphene cluster to oxygen atom. Taking the reference energy of oxygen atom to be -2028.742 eV, obtained by DFT(B3LYP)/3-21G calculation, the adsorption energy of oxygen atom on each H-graphene clusters has been calculated. On increasing the cluster size, the adsorption energy of oxygen atom increases and tends toward saturation from the cluster $C_{70}H_{22}$ as shown in Fig. 12. For the cluster $C_{70}H_{22}$ the bonding energy of adsorbed oxygen atom is 4.011 eV, which is comparable to the corresponding previously reported values 4.80 eV [34] on 4x4 unit cell and 4.42 eV [35] on 3x3 unit cell of graphene. Since the adsorption of oxygen atom on the cluster $C_{70}H_{22}$ almost shows the bulk nature, in the present work, we did not go beyond $C_{70}H_{22}$. All the above discussed parameters have been presented in the Table 1. In the Table 1, *h* is the distance of adatom from the H-graphene basal plane and is defined to be the difference between the Z-coordinates of the adatom and the average of Z-coordinates of the H-graphene sheet. The DOS spectrum of oxygen adsorbed on the H-graphene clusters are as shown in Fig.13, showing that with increasing the cluster size, the Fermi level energy increases and HOMO and LUMO energy levels becomes closer giving small energy gap for large cluster as compared to the smaller cluster of H-graphene.

Table 1. The results of DFT(B3LYP)/3-21G calculations for oxygen atom adsorption on H-graphene clusters considered in the present work

| Cluster | $E_B$ (eV) | $h$ (Å) | Dipole moment (*Debye*) | Charge on O- atom |e| | H-L gap (eV) |
|---|---|---|---|---|---|
| $C_{24}H_{12}$ | 3.581 | 1.90 | 2.329 | -0.393 | 3.169 |
| $C_{30}H_{14}$ | 3.678 | 1.93 | 2.291 | -0.389 | 2.143 |
| $C_{48}H_{18}$ | 3.926 | 1.96 | 2.294 | -0.389 | 1.134 |
| $C_{70}H_{22}$ | 4.011 | 1.97 | 2.356 | -0.389 | 1.741 |

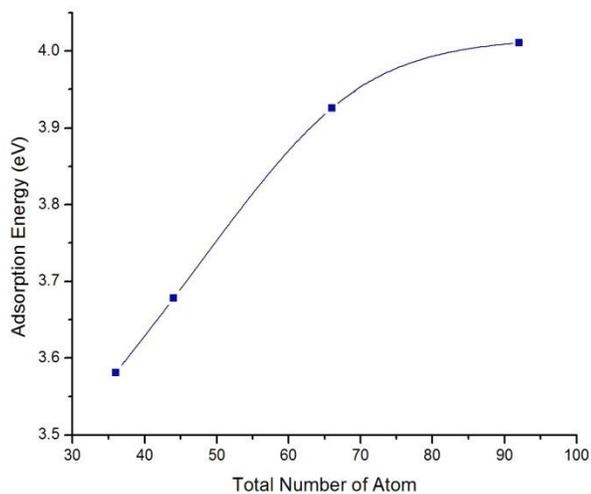

Fig. 12 The variation of adsorption energy of oxygen on H-graphene sheets as a function of corresponding total number of atoms obtained with DFT(B3LYP) levels of calculation using the basis set 3-21G.

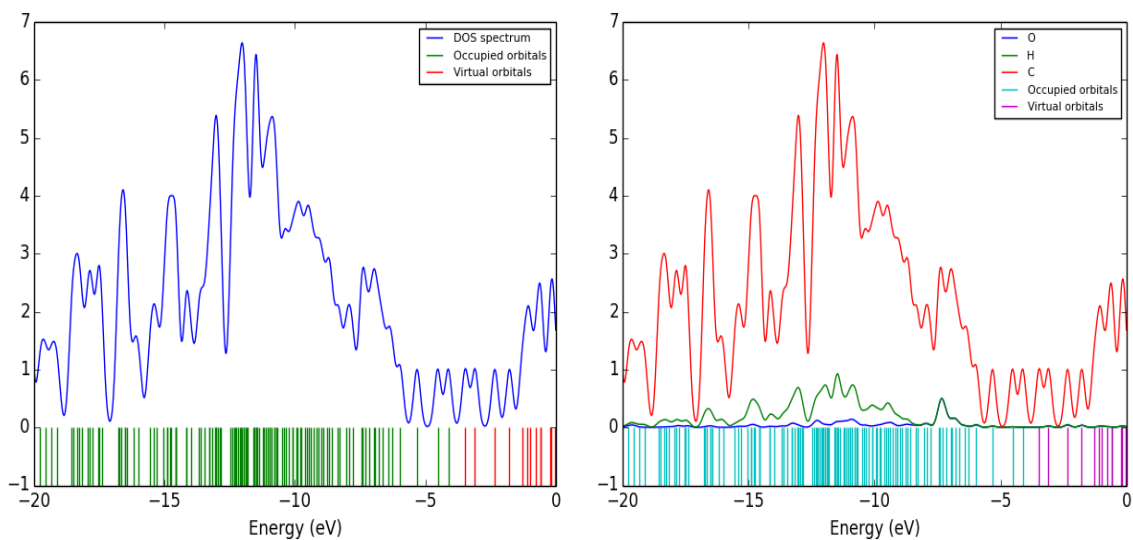

Fig. 13 Total DOS spectrum (left) and partial DOS spectrum (right) of oxygen atom adsorbedon $C_{70}H_{22}$ H-graphene cluster using Mulliken population analysis with DFT(B3LYP)/3-21G levels of calculation.

**3.3.2 Both-sided adsorption**

We have also extended the study of O adsorption on both sides of H-graphene. We first study the interaction of O-pairs with $C_{70}H_{22}$ H-graphene cluster for the case where the first O atom is positioned on the B site and the second O atom is adsorbed on the opposite B site as shown in Fig. 14(a). The adsorption energy per O-atom in this case is found to be 4.083 eV, which is around 2 % greater than that of single O adsorption, which is reasonable because each adatom-O bounds to the same C-atoms in H-graphene leads to the opposite effect to the lattice distortion on H-graphene. Hence, the bond lengths between each adatoms and first and second neighboring C-atoms becomes about 1.48 Å, smaller than that of single O adsorption (1.52 Å). This decrease in bond length increases the covalent-bond interaction, hence, the NBOs charge on each adsorbed O atom is found to be -0.502 |e| which is larger in magnitude as compared to the single O adsorption (-0.389 |e|). The calculated values of dipole moment (0.881 *Debye*) and H-L gap (0.590 eV) in this case are almost one third of that for single O adsorption. The calculated partial DOS spectrum showing the contribution of different atoms in the cluster to the Total DOS is shown in Fig. 15(a).

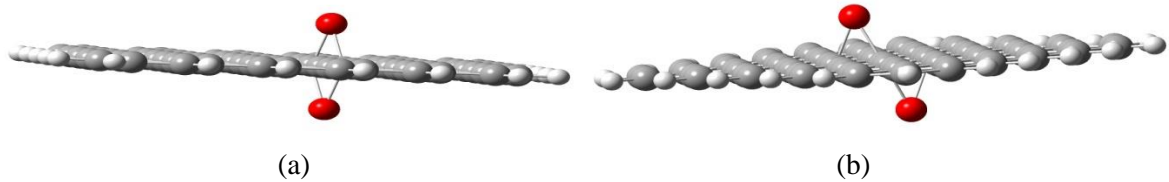

(a)　　　　　　　　　　　　　　　　　(b)

Fig. 14 Schematic views for both side adsorption of O atom on H-graphene cluster $C_{70}H_{22}$ considered in the present work.

The study of both side adsorption has been also extended for the structure as shown in Fig. 14(b). In this case adsorbed O atoms pulls the first neighboring C-atom in H-graphene up while push the second neighboring C-atom down, and thus adsorbed O atom present the consistent effect to the lattice distortion of H-graphene, resulting in stable structure. The adsorption energy per O-atom in this case is found to be 4.652 eV, which is around 14 % greater than that of both side adsorption configuration Fig. 14(a) and around 16 % greater that of single O adsorption. The NBOs charge on adsorbed O atom above the H-graphene sheet is found to be -0.395 |e| and that on O atom below the H-graphene sheet is -0.398 |e|. The calculated values of dipole moment and H-L gap in this case are 0.925 *Debye* and 1.070 eV respectively. The calculated partial DOS spectrum showing the contribution of different atoms in the cluster to the total DOS is shown in Fig. 15(b). All the above discussion is summarized in the Table 2.

Table 2 The results of DFT(B3LYP)/3-21G calculations for both side adsorption of oxygen atom on $C_{70}H_{22}$ with different schemes and comparison of corresponding parameters with single O adsorption.

| Structure | $E_B$ per O-atom (eV) | Dipole moment (*Debye*) | Charge on O-atom (\|e\|) | H-L gap (eV) |
|---|---|---|---|---|
| Fig. 11 | 4.011 | 2.356 | -0.389 | 1.741 |
| Fig. 14(a) | 4.083 | 0.881 | -0.502 [both O-atom] | 0.598 |
| Fig. 14(b) | 4.652 | 0.925 | -0.395 [above sheet] <br> -0.398 [below sheet] | 1.070 |

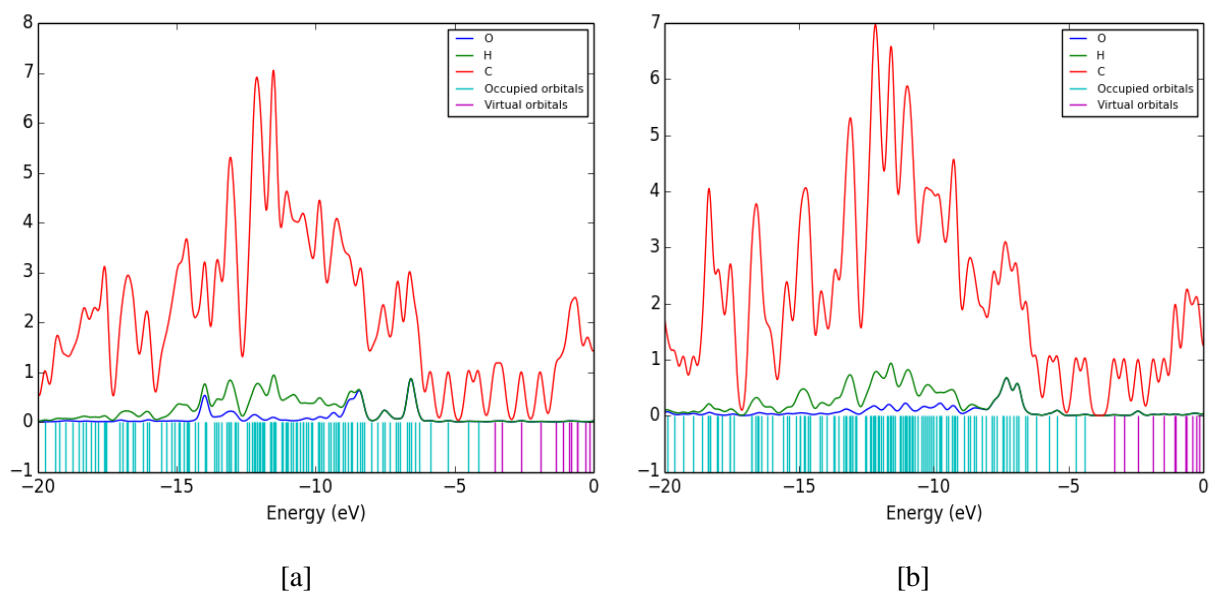

[a]                    [b]

Fig. 15 Partial DOS spectrum for both side adsorption of oxygen atom on $C_{70}H_{22}$ H-graphene cluster using Mulliken population analysis with DFT(B3LYP)/3-21G levels of calculation. [a] Partial DOS for Fig. 14(a) structure [b] Partial DOS for Fig. 14(b) structure.

### 3.3.3 High concentration adsorption

Furthermore, it is of interest to study the relatively higher coverage of O adsorption on H-graphene. With the limit of out computational power, we have just studied the adsorption of three O atoms on H-graphene cluster $C_{70}H_{22}$ as shown in Fig. 16. In this case, the distortion on underlying planner H-graphene sheet is more as compared to single O adsorption and both side adsorptions discussed above. The adsorption energy per O-atom in this case is found to be 4.522 eV, which is greater than that for single O adsorption. The NBOs charge on each adsorbed O atom on H-graphene sheet is found to be -0.376 \|e\|. The calculated values of dipole moment and H-L gap in this case are 5.898 *Debye* and 0.820 eV respectively. The calculated partial DOS spectrum showing the contribution of different atoms in the cluster to the Total DOS is shown in Fig. 17.

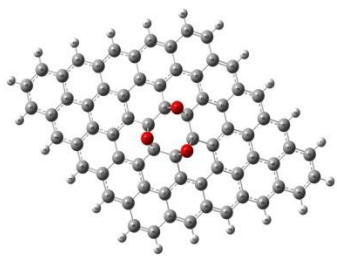

Fig. 16 Optimized structure of $C_{70}H_{22}$ H-graphene cluster with 3 O-atoms adsorbed at alternate B site in the central benzene ring obtained with DFT(B3LYP) levels of calculation using the basis set 3-21G.

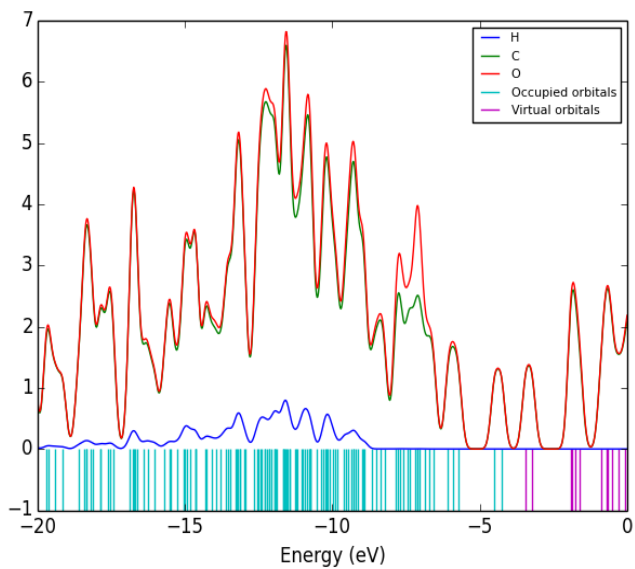

Fig. 17 Partial DOS spectrum for 3 O-atoms adsorbed at alternate B site in the central benzene ring of $C_{70}H_{22}$ H-graphene cluster using Mulliken population analysis with DFT(B3LYP)/3-21G levels of calculation.

## 4. Conclusions and Outlook

We have studied the structural stability of H-graphene as a function of corresponding total number of atom for the graphene sheet having number of carbon atoms from 6 to 160 taking circular and rectangular structures, using first-principles DFT(B3LYP) levels of approximation with the choice of basis set 3-21G implimented by Gaussian 03 set of programs. Our calculation shows that, the relative stability increase with increasing the size of the cluster. From our calculations, we observed that, for rectangular shape clusters, on increasing the cluster size, binding energy per atom ($\delta E/N$) follows perfect linearity with percentage fraction of H-atom as well as C-atom, however, it increases linearly with percentage fraction of C-atom and decreases linearly with percentage fraction of H-atom in the H-graphene clusters. The best fitted lines for $\delta E/N$ (eV) as a function of corresponding percentage fraction of H-atom and C-atom in H-graphene clusters are given as, $\delta E/N = -0.0638 \times$ (% fraction of H-atom) $+ 8.9801$ and $\delta E/N = 0.0638 \times$ (% fraction of C-atom) $+ 2.6004$ respectively. On increasing the cluster size, the H-L gap decreases for both circular and rectangular H-graphene clusters, however, in comparing circular and rectangular cluster containing same number of C-atoms, the H-L gap of former is almost 6 times than that of later. The best fitted line of H-L gap in circular H-graphene clusters with corresponding percentage fraction of H-atoms shows that, the circular H-graphene cluster with about 6 % of H-atoms would have zero H-L gap. Our study on the adsorption of single oxygen atom on H-graphene clusters shows that the adsorption energy increases with increasing the size of the H-graphene cluster and tends toward saturation for relatively larger sized cluster. The charge transfer calculations show that the charge is transferred from oxygen adatom to the H-graphene clusters. Each considered adsorption introduces finite dipole moment in each H-graphene cluster. The adsorption energy per O-atom in case of both side adsorption (one at B site and other at opposite B site below the sheet) is around 2 % greater than that of single O adsorption. The calculated values of dipole moment (0.881 *Debye*) and H-L gap (0.590 eV) in this case are almost one third of that for single O adsorption. The adsorption energy for both side adsorption model such that one at the B site and other at neighboring B site below the H-graphene sheet is around 14 % greater than that of both side adsorption discussed above and 16 % greater that of single O adsorption showing most stable configuration. The calculated values of adsorption energy per O-atom, dipole moment and H-L gap for the case of high coverage adsorption considered in the present work (three O- atoms adsorption on the alternate B site of central benzene ring of H-graphene cluster $C_{70}H_{22}$) shows that this structure is more stable, more polar in nature and softer than the single O adsorbed H-graphene.

**Acknowledgements**

This research was partially supported by the the Abdus Salam International Center for Theoretical Physics (ICTP), Trieste, Italy through the Office of External Activities NET-56.